\documentclass[]{rmaa}

\usepackage{paralist}
\usepackage{psfrag,color}

\def\Liso{\mbox{$L_{\rm iso}$}}
\def\Eiso{\mbox{$E_{\rm iso}$}}
\def\Ep{\mbox{$E_{\rm pk}$}}

\def\Ecol{\mbox{$E_{\gamma}$}}

\def\tbreak{\mbox{$t_{\rm break}$}}

\def\tdur{\mbox{$T_{0.45}$}}
\def\tdurob{\mbox{$T_{0.45}^{\rm obs}$}}
\def\relE{\mbox{\Eiso--\Ep--\tbreak}}
\def\relL{\mbox{\Liso--\Ep--\tdur}}

\def\chs{\mbox{$\chi^2$}}
\def\tj{\mbox{$\theta_{\rm j}$}}
\def\chsr{\mbox{$\chi_r^2$}}

\def\Om{\mbox{$\Omega_{\rm m}$}}
\def\OL{\mbox{$\Omega_{\Lambda}$}}

\def\Ok{\mbox{$\Omega_{k}$}}
\def\Ot{\mbox{$\Omega_{\rm tot}$}}
\def\rde{\mbox{$\rho_{\rm DE}$}}
\def\wz{\mbox{$w(z)$}}
\def\wo{\mbox{$w_0$}}
\def\wu{\mbox{$w_1$}}
\def\wi{\mbox{$w_{\infty}$}}
\def\wp{\mbox{$w_0'$}}
\def\zt{\mbox{$z_t$}}
\def\at{\mbox{$a_t$}}
\def\dl{\mbox{$d_{\rm L}$}}

\def\spose#1{\hbox to 0pt{#1\hss}}
\newcommand\lsim{\mathrel{\spose{\lower 3pt\hbox{$\mathchar"218$}}
     \raise 2.0pt\hbox{$\mathchar"13C$}}}
\newcommand\gsim{\mathrel{\spose{\lower 3pt\hbox{$\mathchar"218$}}
     \raise 2.0pt\hbox{$\mathchar"13E$}}}

\title{Long Gamma-Ray Burst prompt emission properties\\ as a cosmological tool}

\author{C. Firmani,\altaffilmark{1,2}
V. Avila-Reese,\altaffilmark{2}
G. Ghisellini,\altaffilmark{1} 
and G. Ghirlanda\altaffilmark{1}} 

\altaffiltext{1}{INAF-Osservatorio Astronomico di Brera, Italy.}
\altaffiltext{2}{Instituto de Astronom\'ia, Universidad Nacional Aut\'o-noma de M\'exico, Mexico.}

\fulladdresses{
\item V. Avila-Reese: Instituto de Astronom\'ia, Universidad Nacional Aut\'onoma de M\'exico, 
Apdo. Postal 70-264, 04510 M\'exico, D.~F., Mexico (avila@astroscu.unam.mx).
\item C. Firmani: INAF-Osservatorio Astronomico di Brera, via E.~Bianchi 46, I-23807 Merate, Italy and 
Instituto de Astronom\'ia, Universidad Nacional Aut\'onoma de M\'exico, Apdo. Postal 70-264, 04510 
M\'exico, D.~F., Mexico (firmani@merate.mi.astro.it).
\item Giancarlo Ghirlanda and Gabrielle Ghisellini: INAF-Osservatorio Astronomico di Brera, via 
E.~Bianchi 46, I-23807 Merate, Italy (giancarlo.ghirlanda,~gabriele.ghisellini@brera.inaf.it).}

\shortauthor{Firmani et al.} 

\shorttitle{Cosmological constraints with GRBs}

\listofauthors{C.~Firmani, V.~Avila-Reese, G.~Ghisellini, \& G.~Ghirlanda}

\indexauthor{Firmani, C.}
\indexauthor{Avila-Reese, V.}
\indexauthor{Ghisellini, G.}
\indexauthor{Ghirlanda, G.}

\ReceivedDate{2007 January 11} 
\AcceptedDate{2007 February 1} 
\SetYear{2007}

\resumen{Se usa una estrecha correlaci\'on entre 3 propiedades de la emisi\'on $\gamma$ de los 
Estallidos de Rayos Gamma (ERGs) con corrimiento al rojo $z$ conocido \citep{Firmani06} para 
constre\~nir par\'ametros cosmol\'ogicos (PCs) en el diagrama de Hubble (DH) con una muestra de 
19 ERGs en el amplio rango de $z=0.17-4.5$. El problema de la circularidad se resuelve con un 
enfoque bayesiano. Encontramos que la cosmolog\'ia de concordancia $\Lambda$CDM es consistente 
con los datos de los ERGs a nivel de varias pruebas. Si suponemos el modelo $\Lambda$, entonces 
\Om=$0.31^{+0.09}_{-0.08}$ y \OL$=0.80^{+0.20}_{-0.30}$ (1$\sigma$); el caso plano est\'a dentro 
del 1$\sigma$. Suponiendo planitud, obtenemos \Om=$0.29^{+0.08}_{-0.06}$, y fijando \Om=0.28 
obtenemos la ecuaci\'on de estado de la energ\'ia oscura $w=-1.07^{+0.25}_{-0.38}$, estando el 
caso $\Lambda$CDM ($w=-1$) dentro del 1$\sigma$. Dado el bajo n\'umero de ERGs \'utiles no se 
puede a\'un constre\~nir bien la evoluci\'on de $w=w(z)$, pero encontramos que el caso 
$w(z)=-1$ ($\Lambda$CDM) es consistente al 68.3\% CL con los ERGs. Demostramos c\'omo un amplio 
rango de $z'$s en la muestra usada (como es el caso de los ERGs) mejora la determinaci\'on de los 
PCs en el DH. }

\abstract{Recently, a tight correlation among three quantities that characterize the prompt emission 
of long Gamma-Ray Bursts (GRBs) with known redshift $z$, was discovered \citep{Firmani06}. We use 
this correlation to construct the Hubble diagram (HD) with a sample of 19 GRBs in the broad range 
of $z=0.17-4.5$, and carry out a full statistical analysis to constrain cosmological parameters 
(CPs). To optimally solve the problem of circularity, a Bayesian approach is applied. The main result 
is that the concordance $\Lambda$CDM cosmology is fully consistent with the GRB data at the level 
of several tests. If we assume the $\Lambda$ cosmology, then we find \Om=$0.31^{+0.09}_{-0.08}$ and 
\OL$=0.80^{+0.20}_{-0.30}$ (1$\sigma$); the flat-geometry case is within 1$\sigma$. Assuming flatness, 
we find \Om=$0.29^{+0.08}_{-0.06}$, and fixing \Om=0.28, we obtain a dark energy equation of state 
parameter $w=-1.07^{+0.25}_{-0.38}$, i.e. the $\Lambda$CDM model ($w=-1$) is within 1$\sigma$. Given 
the low number of usable GRBs we cannot yet constrain well the possible evolution of $w=w(z)$. However, 
the case $w(z)=-1$ ($\Lambda$CDM) is consistent at the 68.3\% CL with GRBs. It is shown also how a 
broad range of $z'$s in the used sample improves the determination of CPs from the HD, which is the 
case of GRBs as distance indicators.}

\addkeyword{cosmological parameters}  
\addkeyword{cosmology: observations} 
\addkeyword{distance scale}
\addkeyword{gamma rays: bursts}

\begin{document}

\maketitle

\section{Introduction}

The impetuous advance in observational cosmology of the last decade
has prompted new challenges for our understanding of the universe and
its fate, mainly those related to the nature and physics of the dark
energy (hereafter DE) responsible for the current accelerated
expansion of the universe. Stimulated by these challenges, the
frontiers of physics move now in the direction of exploring new
elements of high energy physics, the unification of gravity and
quantum physics, gravity beyond Einstein relativity, and extra
dimensions. At the same time, new astronomical measurements to constrain
DE parameters are being developed with the crucial goal of improving
quality and reducing systematic uncertainties due to astrophysical
effects \citep[e.g.] []{LH05}.  The main task for the new
observational studies is to tell us whether DE can be interpreted in
terms of either a cosmological constant $\Lambda$ (the minimal case)
or something more complex and changing with time, such as scalar
fields.  In this endeavor, alternative and complementary methods and
experiments are mandatory in order to increase the feasibility and rigor of the
results. The use of long gamma-ray bursts (GRBs) as cosmological
distance indicators is gaining popularity as a promising method for
constraining the cosmological parameters related to the dynamics of the
universe. Here we present new advances on this method.

\vspace{0.12cm}

As the most powerful explosions in the universe, long GRBs are of
great interest for observational cosmology because they can be
detected up to very high redshifts, the current record with
spectroscopic determination being GRB 050904 at $z=6.29$
\citep{Kawai05}. Ghirlanda, Ghisellini, \& Lazzati (2004a) have discovered a tight correlation
between the rest frame collimation corrected energy \Ecol\ and the
peak energy \Ep\ of the $\nu F_\nu$ prompt emission spectrum for a
sample of GRBs with known $z$. The use of this correlation has proved
to be very useful as a method for ``standardizing'' the GRB energetics
and its further application for constructing the Hubble diagram.

\vspace{0.12cm}

The ``Ghirlanda'' relation has been already used to obtain cosmological 
constraints, after applying adequate approaches to overcome the problem 
that, due to the lack of a local (cosmology-independent) calibration, this
relation actually depends on the cosmological parameters that we pretend to 
constrain (Ghirlanda et al. 2004b; Firmani et al. 2005; Xu, Dai, \& Liang 2005; 
Ghirlanda et al. 2006). {\it As a result, the accelerated expansion of the universe 
at the present epoch was confirmed independently with GRBs}. Interestingly 
enough, the marginal inconsistency of the ``gold set'' of Type-Ia supernovae (SNIa
hereafter) with the simple flat-geometry Friedmann-Lema\^itre-Robertson-Walker
(FLRW) cosmology including the cosmological constant ($\Lambda$-cosmology) (e.g. 
Riess et al. 2004; Alam, Sahni, \& Starobinsky 2004; Choudhury \& Padmanabhan 2004; 
Jassal, Bagla, \& Padmanabhan 2005; Nesseris \& Perivolaropoulos 2005b) {\it is 
eliminated when the GRB data are added} \citep{fgga05,Ghirla05}.  

It is important to stress that GRBs (i) are detected from redshifts much 
higher than SNIa, and (ii) some degeneracies in determining the cosmological 
parameters are reduced if the observational sample  displays a broad range in 
redshifts, attaining high values of $z$ \citep[e.g.][]{weller02, LH03, nesseris04, 
Ghisellini05}.  In \S 4 this question will be amply discussed, showing by concrete 
examples {\it why a sample broad in redshifts improves the determination of the 
cosmological parameters.}

The \Ecol--\Ep\ relation takes into account the GRB collimation--corrected energy, 
\Ecol=\Eiso(1-$\cos\tj$), where \Eiso\ is the isotropic--equivalent energy and \tj\ 
is the semi-aperture jet angle. The determination of this angle is model dependent. 
For the uniform jet model in the standard fireball scenario, \tj\ can be determined 
by the time \tbreak when the afterglow light-curve becomes steeper. For a homogeneous 
circumburst medium $\tj\propto t_{\rm break}^{3/8}$ (e.g. Sari, Piran, \& Halpern 
1999), while for a wind circumburst density profile decreasing as $r^{-2}$, 
$\tj\propto t_{\rm break}^{1/4}$ \citep[]{Nava06}. Note that to estimate the jet angle 
from $t_{\rm break}$ one must also assume a specific value of the density of the 
circumburst material, and the efficiency to convert the fireball kinetic energy into 
the radiation emitted during the prompt phase. Liang \& Zhang (2006; see also Firmani 
et al. 2006a) found a purely empirical multi-variable correlation among \Eiso, \Ep\ 
and \tbreak\ (which is then model-independent and assumption-free). They used this 
correlation to constrain the cosmological parameters \citep[see also][]{Xu05,Ghirla05}.

In \citet[][hereafter Paper I]{Firmani06} we have searched for empirical correlations 
among $\gamma-$ray prompt quantities alone. In the GRB rest frame the considered quantities 
were the bolometric corrected \Liso\ and \Eiso, the spectral peak energy \Ep, and the 
light-curve variability $V$ and duration \tdur\ \citep[as defined in][for more details 
see \S~2 below]{Reichart01}. In Paper I a $\Lambda$CDM cosmology with \Om=0.3, \OL=0.7, 
$h$=0.7 was assumed to calculate luminosity distances. For the sample of 19 GRBs, for 
which all the above quantities can be defined, we have found a very tight multi-variable 
correlation among three quantities, namely $\Liso\propto E_{\rm pk}^{1.62} T_{0.45}^{-0.49}$. 
Within the framework of the fireball scenario, the tightness of the correlation is 
explained by its scalar nature. We have also estimated the correlation among \Eiso, \Ep\ 
and \tbreak\ \citep{liang05} for the 15 GRBs with measured \tbreak\ of our sample, and 
have proved that the \relL\ relation is as tight as the \relE\ one.

Similarly to the ``Ghirlanda'' (or ``Liang \& Zhang'') correlation, the \relL\ correlation 
can be used as a cosmic ruler for cosmographic purposes. From a practical point of view, 
the great advantage of the \relL\ correlation is that {\it it involves quantities related 
only to the $\gamma-$ray prompt emission}. Thus, the establishment of this correlation avoids 
the need to monitor the afterglow light-curve in order to derive \tbreak\ which enters both 
in the ``Ghirlanda'' and in the ``Liang \& Zhang'' correlation. 

In this paper we analyze in detail the cosmographic application of the \relL~ relation by 
using the current dataset. The Hubble diagram is constructed up to redshifts as high as $z=4.5$. 
We also describe the Bayesian formalism to solve the `circularity problem' and compare it 
with other formalisms. Note that this problem, at least formally, is also present for SNIa 
samples, as is the case for the recent SN Legacy Survey \citep{Astier06}. Thus, the Bayesian 
formalism can also be used to obtain improved cosmological constraints from SNIa samples.

The GRB sample and the \relL\ correlation are presented in \S~2. The changes of the correlation 
with cosmology are analyzed in \S~3, where we test the robustness of such a correlation for 
cosmographic purposes. In \S~4 we present our approach to parametrize the evolution of DE, and 
we discuss the degeneracies present in the set of dynamical cosmological parameters We also 
discuss the Bayesian formalism for solving the circularity problem, comparing it with the 
conventional $\chi^2$ approach. In \S~5 we present the constraints on the parameters that 
describe the geometry and dynamical evolution of the Universe obtained with the sample of 
19 GRBs. The summary and a brief discussion on the current shortcomings and the future of 
the method presented here are given in \S~6.


\section{The sample and the \relL\ relation}

The sample of GRBs with known redshifts and with the necessary observational information 
available was presented in Paper I. The {\it rest frame} \relL\ correlation presented in 
Paper I involves:
\begin{itemize} 
\item the bolometric corrected isotropic energy $E_{\rm iso}$, computed in the rest frame 
$1-10^4$ keV energy range;
\item the peak energy \Ep\ of the $\nu F_{\nu}$ prompt emission time integrated spectrum;
\item  the time  \tdur\  spanned by  the  brightest 45\%  of the  total light curve counts 
above the background and calculated in the 50-300 keV rest frame energy range\footnote{We 
used the recipe proposed by \citet{Reichart01} to transform the  observed energy range to 
the rest frame, and the time binning of {\it HETE--II}, 164-ms (see Paper I).}.
\end{itemize}

In addition to the spectroscopically measured redshift $z$, the observational data required 
to estimate \Liso, \Ep\ and \tdur\ are the peak flux $P$, the fluence $F$, the spectral 
parameters of a given spectral model (in most cases the Band et al. 1993 model) and the light 
curve (to estimate \tdurob).  The uncertainties in these observables are appropriately 
propagated to the composite quantities \Liso, \Ep\ and \tdur\ under the assumption of no 
correlation among the measured errors. Note that all the above quantities (except $z$) are 
obtained exclusively from the $\gamma$-ray prompt emission of the burst.

In Paper I we have used a flat-geometry $\Lambda$ cosmological model with \Om=0.3, \OL=0.7, 
and $h$=0.7 to calculate the GRB luminosity distances, \dl, and to estimate \Liso. Then, for 
the 19 GRBs with available observational data a multi-variable regression analysis, {\it taking 
into account errors in all the variables}, provided the following best fit:
\begin{eqnarray} 
\Liso &=& 10^{52.11\pm0.03} 
\left({\Ep \over 10^{2.37}{\rm keV}}\right)^{1.62\pm0.08}
\nonumber \\
&~&
\left({\tdur \over 10^{0.46}{\rm s}}\right)^{-0.49\pm0.07} \,
{\rm erg~s^{-1}}
\label{Liso} 
\end{eqnarray}
For a detailed discussion of this correlation, the error estimates, the comparison with other 
correlations and its interpretation we refer the reader to Paper I.


\begin{figure}[!t]
\centerline{\includegraphics[width=0.98\columnwidth]{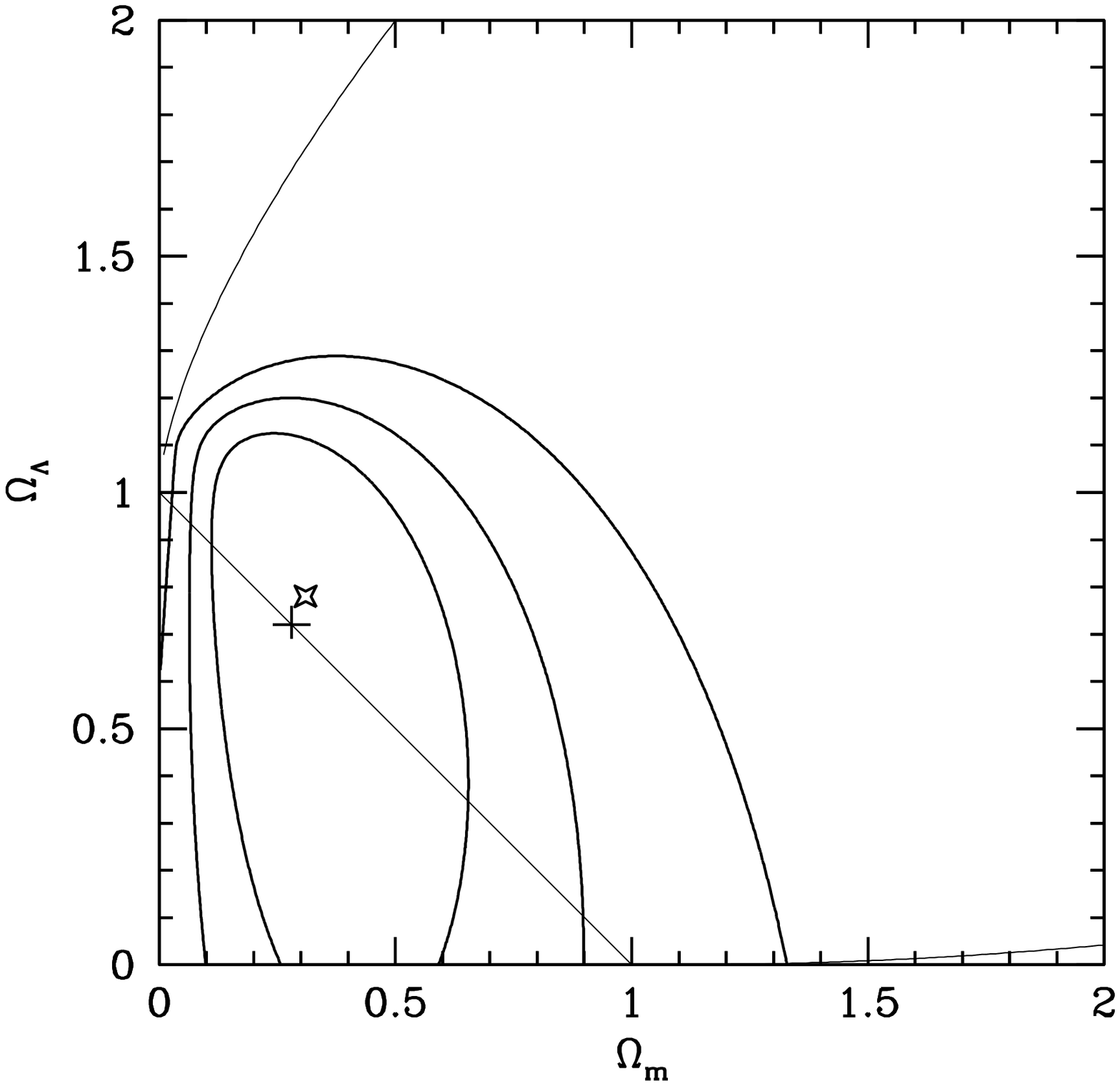}}
\caption{Contours at 68.3\%, 95.5\%, and 99.7\% CL's obtained by projecting on the 
(\Om, \OL) plane the \relL\ relation $\chsr$ derived from the fit of the GRB data 
at each value of the (\Om, \OL) pair. The star shows where the $\chsr$ reaches its 
minimum, while the cross indicates the concordance cosmology. This plot shows that 
the relation \relL\ is sensitive to cosmology, so that it may be used to discriminate 
cosmological parameters if an optimal method to circumvent the circularity problem 
is used. The diagonal line corresponds to the flat geometry cosmology, the upper curve 
is the loitering limit between Big Bang and No Big Bang models, and the lower curve 
indicates the division between accelerating and non-accelerating universes.} 
\label{fig1}
\vspace{0.3cm}
\end{figure}


\section{A cosmological test for the \relL\ correlation}

A preliminary cosmological test concerns the sensitivity of the \relL\ correlation to the 
dynamical cosmological parameters. In Paper I we have assumed the currently conventional 
cosmological model. Now, we will analyze how the correlation and its scatter change from 
one cosmology to another. For each $\Lambda$ FLRW cosmology characterized by (\Om, \OL) we 
perform the multiple variable regression analysis on the dataset, using the same method 
described in Paper I. In this way, the (best fit) \relL\ correlation, its relative scatter 
and the corresponding $\chsr$ value for each cosmology are obtained and can be used to 
assign a probability to the (\Om, \OL)-pair \citep{gglf04}.


\begin{figure}[!t]
\centerline{\includegraphics[width=0.98\columnwidth]{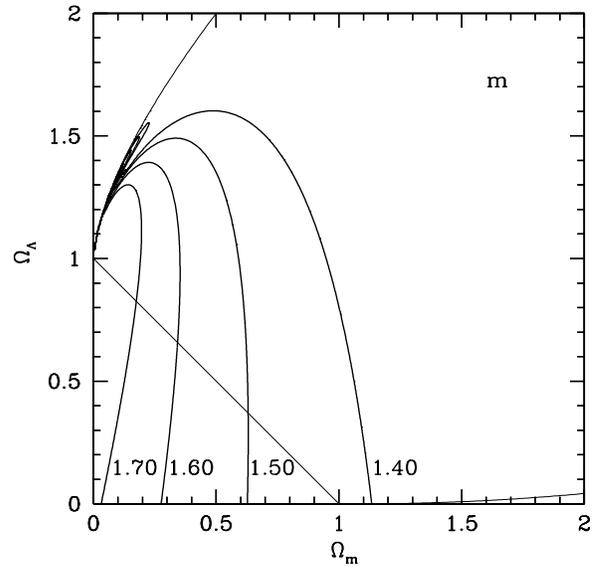}}
\caption{Contours of constant $m$ in the (\Om, \OL) plane. $m$ is the 
power of \Ep\ in the \relL\ relation: $\Liso\propto E_{\rm pk}^m T_{0.45}^{-n}$. 
The other curves in the plot are as in Figure~1.}
\label{fig2}
\vspace{0.4cm}
\end{figure}


In Figure \ref{fig1} we show the resulting contours at 68.3\%, 95.5\%, and 99.7\% confidence 
levels (hereafther CL's), which measure how the $\chsr$ (related to the scatter) of the \relL\ 
relation changes with the cosmological parameters. Figure~\ref{fig1} reveals an important 
sensitivity of the scatter on cosmology and shows the rather surprising result that {\it the 
smallest scatter is found for (\Om, \OL) = (0.31, 0.78), close to the concordance model 
(\Om, \OL) = (0.28, 0.72) which falls deep inside the 68.3\% confidence level region}. This 
simple and direct (`scatter-scanning') formalism for constraining cosmological parameters does 
not optimize the use of the available information and is particularly sensitive to the loitering 
line singularity \citep{fgga05}. However, it already shows the potentiality of the \relL\ 
relation for cosmographic purposes. This encourages us to use a more sophisticated formalism 
in order to obtain more accurate cosmological constraints (see \S~4.3).

\begin{figure}[!t]
\vspace{0.1cm}
\centerline{\includegraphics[width=0.98\columnwidth]{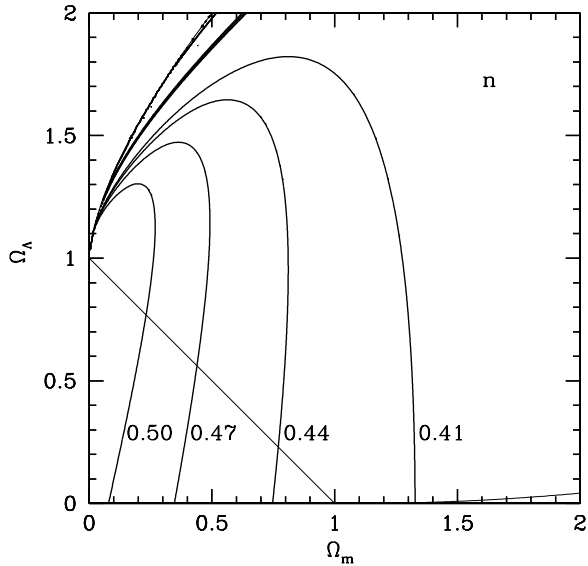}}
\caption{Contours of constant $n$ in the (\Om, \OL) plane. $n$ is the power 
of \tdur\ in the \relL\ relation: $\Liso\propto E_{\rm pk}^m T_{0.45}^{-n}$. 
The other curves in the plot are as in Figure~1.}
\label{fig3}
\vspace{0.3cm}
\end{figure}

Figures~\ref{fig2} and \ref{fig3} illustrate, respectively, how the powers $m$ and $n$ of the 
\relL\ relation ($\Liso\propto E_{\rm p}^m T_{0.45}^{-n}$) change, in the (\Om, \OL) plane. The 
lines in Figures~\ref{fig2} and \ref{fig3} are not to be confused with CL contours on the 
cosmological parameters. Notice the behavior of the isocontours near the loitering curve, where 
the dependence of \dl\ on the cosmological parameters becomes singular. The exponents of the \relL\ 
relation do not change dramatically in a wide range of (\Om, \OL) values, even if these changes 
are significantly larger than the small standard deviations of the exponents obtained in the fits 
(see e.g. Eq.~\ref{Liso}). We hope that these results can help for the theoretical interpretation 
of the obtained correlation, indicating the(rather small) range of the allowed $m$ and $n$ values.


\section{Constraining the cosmological parameters in the Hubble diagram}

\subsection{Cosmological Models with Dark Energy}

The accelerated expansion of the universe is often explained by the dominance in the present-day 
universe of a self-repulsive medium (DE) with an equation of state parameter 
$w = p_{\rm DE}/\rde c^2 < -1/3$. The simplest interpretation of DE is the homogeneous and inert 
cosmological constant $\Lambda$, with $w=-1$ and $\rde=\rho_{\Lambda}=$const. The combinations of 
different cosmological measurements tend to favor models where DE is $\Lambda$ and 
(\Om, \OL)$\approx$(0.28, 0.72) \citep[the so-called concordance model, e.g.][]{Spergel03,Tegmark04,
Seljak05}. Nevertheless, it is important to note that, due to a variety of degeneracies in the parameter 
space, there are not yet any reliable joint constraints to the complete set of cosmological parameters, 
even after combining different cosmological probes and data samples \citep[e.g.][]{Bridle03}. Different 
probes can even lead to constraints which are not in complete agreement among them (when treated 
separately), as is the case of \textit{WMAP} observations of the CMB and the ``gold set'' of Type-Ia 
SNe (Jassal et al. 2005). Note that a more recent analysis based on the SN Legacy Survey has reduced 
this apparent discrepancy by favoring the simple flat $\Lambda$ cosmological model \citep{Astier06}.

Through samples of ``standard candle'' objects, such as Type-Ia SNe or GRBs, it is possible to 
construct the Hubble diagram and, by comparing the data points with the model curves (for different 
choices of \Om\ and \OL), to constrain these cosmological parameters. It is clear that, allowing $w$ 
to have values different from $-1$, or even evolving, increases the number of free parameters to fit. 
Up to now, the existing datasets do not allow to fit together all these parameters. The most common 
approach is to fit only a couple of cosmological parameters, keeping all the others fixed. Such an 
exercise is in any case important since, for instance, the cosmological constant explanation of DE 
faces serious theoretical problems \citep[see for reviews e.g.][]{Padma03,Sahni04}. Therefore, 
alternative scenarios, where $w$ is different from $-1$ or even variable with $z$, have been proposed 
and extensively investigated.

According to the approach mentioned above, we proceed here in three stages. First, we constrain 
the two parameters, (\Om, \OL), of the (minimal) $\Lambda$ cosmology ($w=-1$), and further check 
whether the concordance model (implying flat geometry) is statistically consistent with the 
constraints.  Next, we generalize to models with $w=$const (static DE), but assuming a flat 
geometry in order to have only 2 fitting parameters, \Om\ and $w=$const. Finally, we generalize 
to evolving (dynamical) DE models, where $w$ changes with $z$ according to a parametric form, 
assuming a flat geometry and \Om\ = 0.28. In the two last stages, with some redundancies, we again 
check whether the concordance model is statistically consistent with the constraints, i.e. whether 
$w=-1=$ const is within the 68.3\% CL region. For the dynamical DE models, we explore also how much 
the observational constraints favor the case of an evolving or a static $w$. Note that any 
parametrization of $w(z)$ is limited and arbitrary.

To model an evolving DE we use a rather general parametrization for $w$ proposed by Rapetti, Allen, 
\& Weller (2005):
\begin{equation} 
\wz = \wo + \wu \frac{z} {\zt + z},
\label{w} 
\end{equation}
where the parameter $\wo$ gives the present-day (i.e. at $z=0$) equation of state; $\wu = \wi - \wo$ 
gives the increment of $w$ from the present value to $z=\infty$ and $z_t$ is a redshift transition 
scale. Note that $z_t$ should not be confused with the transition redshift where the expansion goes 
from decelerating to accelerating). The derivative of $\wz$ at present is $\wp = \wu / \zt$. The 
evolution of the Hubble parameter is given by
\begin{equation} 
H^2 = H_0^2 \left[ \Om (1+z)^3 + \OL f(z) + \Ok (1+z)^2 \right]
\label{H} 
\end{equation}
where $\Ok = 1 - \Om - \Omega_{\uppercase{de}}$,
\begin{equation} 
f(z) = (1+z)^{3 (1+\wi)} e^{-3 (\wi - \wo) g} \, ,
\label{f} 
\end{equation}
and
\begin{equation} 
g = \frac{1-\at} {1-2 \at} \ln \left[ \frac{1-\at} {a(1-2\at)+\at} \right] \, ,
\label{g} 
\end{equation}
where $a$ is the scale factor and $\at=1/(1+\zt)$ is the corresponding transition scale factor.

The simple linear approximation commonly used in previous works \citep[e.g.][]{Riess04, fgga05}, 
is obtained by making $\zt$ arbitrarily large and assigning a given value to $\wp$, which in this 
case is the slope of the $w(z)$ function. The parametrization of \citet{linder03} is recovered by 
setting $\zt = 1$ ($\at=1/2$). Figure~4 shows the family of parametric curves given by Eq.~(\ref{w}).  
Here $\wo=-1$, $\wu = -1$, $0$, $1$ and $2$ and $\zt =0.5$ (short-dashed), $1.0$ (long-dashed) and 
$1.5$ (point-dashed). The dotted lines show the linear approximation at present.


\begin{figure}[!t]
\vspace{0.1cm}
\centerline{\includegraphics[width=0.98\columnwidth]{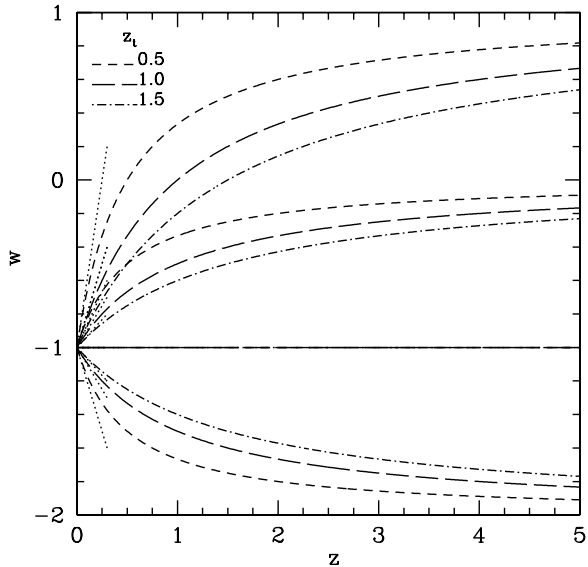}}
\caption{ Dependence of the equation of state parameter $w$ on $z$ as 
described by Eq.~(\ref{w}). For the plot, \wo=$-1$ is assumed. Values 
of \wu=2, 1, 0 and $-1$ (from top to bottom respectively) and of \zt=0.5, 
1.0, and 1.5 (see labels inside the panel) for each \wu\ are used. The 
dotted lines are the tangent lines of each curve at $z=0$ and represent 
the linear approximation.}
\label{fig4}
\vspace{0.4cm}
\end{figure}


\vspace{0.12cm}

Three aspects related to the task of constraining $w(z)$ are worth of mention. 

\begin{enumerate}
\item Methods based on the construction of the Hubble diagram with a given class of standard candles 
provide the primary source of information on the evolution of DE \footnote{We notice that besides of 
the methods based on the luminosity distance--$z$ diagram are also the methods based on the angular 
diameter distance--$z$ diagram (e.g. the baryonic acoustic oscillations).}, which is expected to 
become dominant only at low redshifts ($\lsim 1$).
\item If the redshift range of the sources is small, in particular limited to low $z'$s, then the DE 
parameters and evolution can be constrained only in a limited way (see \S~4.2).
\item The constraints on $w(z)$ depend on the (arbitrary) assumed parametrization for $w(z)$. In fact, 
the space of all possible parametrizations is infinite-dimensional. By choosing ``reasonable'' 
parametrizations, both in the physical sense and in that of the limitations of the observational data, 
the main information we may intend to derive refers only to general aspects such as whether there is 
evidence or not of DE evolution, and what is the direction of this evolution \citep[e.g.][]{LH05}. 
Adequate parametrizations are those with a minimum number of parameters but allowing the widest range 
of variation of $w$ over the $z$ range in which $w\prime(z)$ is best constrained by the given class 
of standard candles \citep{Amol05}. For the parametrization of Eq. (\ref{w}), the smaller is \zt, the 
larger the allowed change for $w(z)$ at low $z'$s, where observational data are available (see 
Figure~\ref{fig4}).
\end{enumerate}

\vspace{0.15cm}

\subsection{The Hubble Diagram for High Redshift Objects}

We now discuss some aspects related to the Hubble diagram used to constrain the cosmological 
parameters. GRBs are the natural objects for extending cosmographic studies up to very high 
redshifts, and thus for inferring the behavior of DE, in particular, whether and how it evolves. 
A fundamental issue is, therefore, to understand all the power of the information which can 
be extracted from using the GRBs as standard candles extending up to very high redshifts. In 
particular, one should be aware of the several degeneracies (correlations) that appear among 
the cosmological parameters at different redfshifts.

\begin{figure}[!t]
\vspace{0.1cm}
\centerline{\includegraphics[width=0.98\columnwidth]{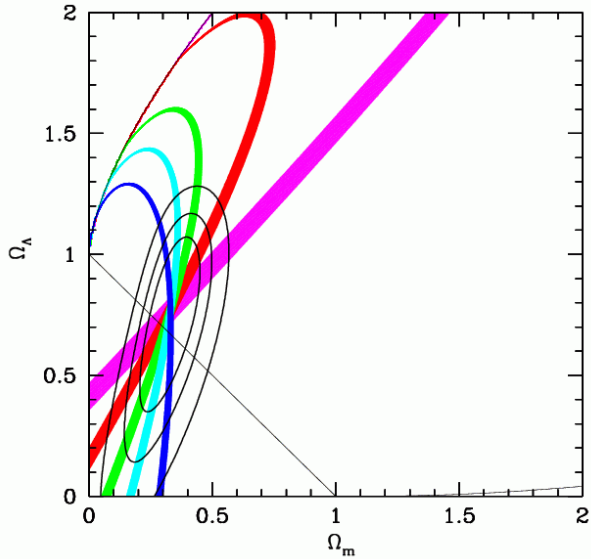}}
\caption{Regions of $\pm 1\%$ variation around lines of constant \dl\ in the 
(\Om, \OL) plane, assuming that each line passes through the fiducial point 
(\Om,\OL) = (0.33, 0.77). In counterclockwise rotation, the regions are at 
redshifts 0.5, 1.0, 1.5, 2.0, and 3.0, respectively [in the electronic version 
of this paper $z=0.5$ (magenta), 1.0 (red), 1.5 (green), 2.0 (cyan), and 3.0 
(blue)]. This plot illustrates the degeneracy between the parameters \Om\ and 
\OL\ ($\Lambda$ cosmology) and how this degeneracy does change with $z$. The 
ellipses are contours at 68.3\%, 95.5\%, and 99.7\% CL's for the fit to a $\Lambda$ 
cosmology from the GRB Hubble diagram, using a \relL\ relation supposed to be 
known, and therefore fixed and cosmology-independent (see text for more details). 
Note that the main orientation of the ellipses is along the ``stripes'' with 
$z \approx 1.5$, which corresponds roughly to the typical redshifts of the GRB 
sample. The other curves in the plot are as in Figure~1.
\label{fig5}}
\label{OOdlines}
\vspace{0.3cm}
\end{figure}

To study such degeneracies and to understand the shape of the CL's in the parameter space, it 
is instructive to explore the behavior of the luminosity distance \dl\ at different redshifts 
$z$ in a given cosmological parameter space. Consider first the $\Lambda$ cosmology, where the 
parameters are \Om\ and \OL. In counterclockwise rotation, the stripes shown in Figure~\ref{fig5} 
represent the regions of the (\Om, \OL) plane where \dl varies by $\pm$ 1\% for z = 0.5, 1, 1.5, 
and 3, respectively, assuming that each stripe passes through the fiducial point (\Om, \OL)=(0.33, 0.77) 
(see below the reasons for this choice).


The stripes in Figure \ref{fig5} show that the degeneracy (correlation) between \Om\ and \OL\ 
varies with $z$. This has immediate implications for cosmographic methods based on luminosity 
distance measurements. Taking into account the measurement uncertainties, a specific ``standard 
candle'' determines a range of luminosity distances \dl\ and consequently a stripe on the 
(\Om, \OL) diagram.  For a sample of standard candles characterized by a small range in redshifts, 
the corresponding CL's in the (\Om, \OL) diagram will be very elongated (high degeneracy) and 
will have the major axis oriented in the direction of the stripe of the average redshift of the 
sample. Therefore, a counterclockwise rotation of the CL's is expected when the average redshift 
of the standard candle sample used to derive the CL's increases.  This easily explains why the 
CL's derived by using SNIa data are elongated and oriented approximately along the direction of 
the $z\sim 0.6$ stripe (see Figure \ref{fig8}), while the contours derived using our GRB sample, 
of larger average redshift ($z\sim 1.5$) are more ``vertical''.  Note that, although our GRB 
sample contains a factor of 10 fewer objects than SNIa, it produces a comparatively narrow contour 
region, thanks to the broad distribution of redshifts of the GRBs in the sample.

\vspace{0.17cm}

Figure~\ref{fig5} also shows that the width of the stripes (i.e. the uncertainty in (\Om, \OL) 
associated to a given luminosity distance) decreases for larger $z'$s. This is a consequence of 
the topology of the surfaces of constant \dl: at low redshift the surface is a gently tilted 
plane, at high redshifts the surface is more warped, and there appears a ``mountain'' with a peak 
close to \Om$\sim 0.0$ and \OL$\sim 1$. As a consequence, the stripes at high $z'$s are curved, 
and at very high $z'$s they surround the ``mountain peak''. Note that, as a consequence of the 
increasing slope of the \dl\ surface, the width of the stripes at high redshifts becomes narrower 
for large \OL--values.

\vspace{0.17cm}

From Figure~\ref{fig5} we conclude that in order to reduce the degeneracy and improve the accuracy 
of the constraints of \Om\ and \OL\ by using the luminosity distance method, the sample of observed 
sources should span a range of redshifts as large as possible.


\begin{figure}[!t]
\vspace{0.05cm}
\centerline{\includegraphics[width=0.98\columnwidth]{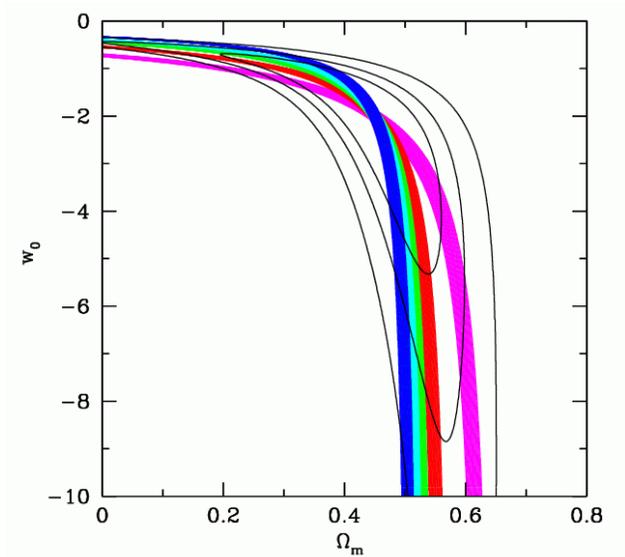}}
\caption{ Same as Figure \ref{fig5} but in the (\Om, $w$) plane for a 
flat cosmology with static DE. The fiducial point is (\Om, $w$) = (0.45, 
$-$2.00).  Taking the vertical regions of the stripes, the redshifts are 
0.5, 1.0, 1.5, 2.0, and 3.0 from  right to left, respectively [in the 
electronic version of this paper $z=0.5$ (magenta), 1.0 (red), 1.5 (green), 
2.0 (cyan), and 3.0 (blue)].  For low values of \Om, $w$ is almost independent 
from \Om, while the opposite happens for high values of \Om. The dependence 
of the \Om--$w$ degeneracy on $z$ is weak. As in Figure~\ref{fig5}, the bent 
ellipses are the contours of CL from the corresponding GRB Hubble diagram, 
using a \relL\ relation supposed to be known, and therefore fixed and cosmology-independent.}
\label{fig6}
\vspace{0.3cm}
\end{figure}



\begin{figure}[!t]
\centerline{\includegraphics[width=0.96\columnwidth]{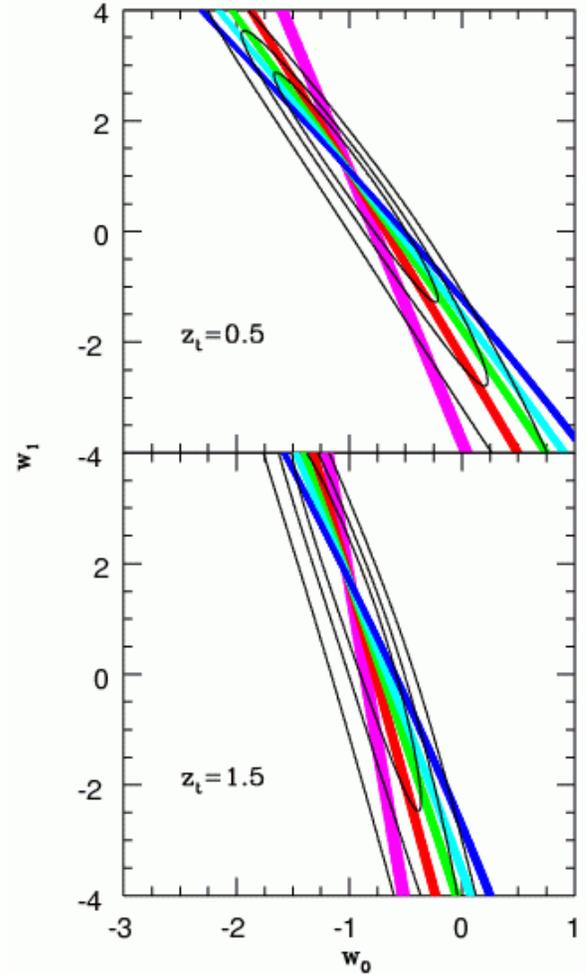}} 
\caption{Same as Figure~\ref{fig5} but in the (\wo, \wu) plane for a flat 
geometry cosmology with dynamic DE and \Om = 0.28. The evolving $w(z)$ is 
parametrized according to Eq.~(\ref{w}) with \zt =0.5 (upper panel) and 
\zt=1.5 (lower panel). The fiducial points are (\wo, \wu) = ($-$1.00, 1.08) 
and (\wo, \wu) = ($-$1.01, 0.61), respectively. Taking the lower part of the 
stripes, the redshifts are 0.5, 1.0. 1.5, 2.0 and 3.0 from left to right, 
respectively [in the electronic version of this paper $z=0.5$ (magenta), 
1.0 (red), 1.5 (green), 2.0 (cyan), and 3.0 (blue)]. As in Figure~\ref{fig5}, 
the ellipses are the contours of CL from the corresponding GRB Hubble diagram, 
using a \relL\ relation supposed to be known, and therefore fixed and cosmology-independent.}
\label{fig7}
\vspace{0.4cm}
\end{figure}


\vspace{0.17cm}

The fiducial point, (\Om, \OL) = (0.33, 0.77), and the CL contours in Figure~\ref{fig5} have been 
calculated in the following way. We began by using arbitrary trial values for \Om\ and \OL\ to 
define a fiducial (unique) \relL\ relation. Further, we calculated the \chsr's in the whole (\Om, \OL) 
plane using such fiducial \relL\ relation to assign the luminous distance to each GRB of known $z$. 
If the minimum of the \chsr's was smaller than the \chsr\ corresponding to the trial (\Om, \OL) values, 
then the (\Om, \OL) values corresponding to the minimum \chsr\ were used to define a new fiducial \relL\ 
relation in a new iterative step. This procedure was repeated until convergence. The CL's correspond 
to the 68.3\%, 95.5\%, and 99.7\% (1$\sigma$, 2$\sigma$ and 3$\sigma$) probabilities provided by the 
$\chi^2$ statistics. The procedure should be considered here as a (naive) simulation to constrain 
\Om\ and \OL\ by using a {\it unique and well calibrated} \relL\ relation. Interestingly enough, the 
convergence (\Om, \OL) values in our exercise lie close to those of the concordance cosmology. However, 
we remark that this procedure is based on incorrect assumptions; it is introduced here only for 
heuristic reasons.

\vspace{0.17cm}

With  an analysis similar to that applied in Figure~1, we also study the behavior of \dl\ in the 
diagrams (\Om, $w$=const) and (\wo, \wu) for flat geometry cosmological models. In the latter case, 
we have further assumed that $\Om = 0.28$ and used Eq.~\ref{w} with $\zt = 0.5$ and $1.5$ to 
describe the evolving DE. Figures~\ref{fig6} and \ref{fig7} show the regions of \dl=const for 
$z = 0.5, 1, 1.5, 2,$ and $3$ (see details in the figure captions) assuming in each case that the 
center of each stripe passes through a given fiducial point.
 
\vspace{0.05cm}

The fiducial point in each case is [(\Om, $w=0.45, -2.00$), (\wo, \wu$=-1.00, 1.08$) for \zt=0.5, 
and (\wo, \wu$=-1.01, 1.61$) for \zt=1.5]. The CL's in the (\Om, $w$) and (\wo, \wu) plane, 
represented in Figure~\ref{fig6} and Figure~\ref{fig7} respectively, were computed following the 
same procedure described above for the (\Om, \OL) plane. The 1$\sigma$, 2$\sigma$, and 3$\sigma$ 
CL's in Figure~\ref{fig6} and \ref{fig7} were provided by the corresponding $\chi^2$ statistics. 
Figures~\ref{fig6} and \ref{fig7} show the degeneracies between $w=$const and \Om, and between 
\wo\ and \wu\ (for \zt=0.5 and 1.5), and how these degeneracies depend on $z$.

\vspace{0.05cm}

To summarize: the study of the $d_{\rm L}(z)$ surfaces in the different planes helps us to understand 
the orientations of the CL regions for different samples of cosmological probes, characterized by 
different average redshifts. This study makes intuitively clear the need to have probes distributed 
in a large range of redshifts. This in turns implies that SNIa and GRBs complement each other in a 
natural way.

\subsection{The Bayesian Formalism}

Now we will explore how the correlation \relL\ can be used to constrain cosmological parameters 
through the Hubble diagram. In the previous section we introduced the concept of a {\it unique
well calibrated} \relL\ relation; however, this is not the present case. In fact the \relL\ 
{\it depends on the assumed cosmology}. Therefore the crucial issue in this undertaking is what
has been called the ``circularity problem'': we attempt to constrain the cosmological parameters 
using a correlation which is cosmology-dependent. This problem arises because, due to the lack 
of detected low$-z$ GBRs, the \relL\ correlation can not be calibrated at low redshifts, where 
the flux is not affected by a specific cosmology. Another way to calibrate this kind of 
correlation is with a sample of high-redshift GRBs in a considerably small redshift bin. \citet{Ghirla05} 
have calculated that $\sim 12$ GRBs with $z \in (0.9,1.1)$ can be used to calibrate the 
``Ghirlanda'' relation with a precision higher than $1\%$. This number might be reached in a 
few years of observations mainly due to the fact that the jet break time measurement (which 
enters in the ``Ghirlanda'' correlation) requires a time-consuming follow up campaign of the GRB 
optical/NIR afterglow. We estimate that a similar number of GRBs might also be used to calibrate 
the \relL\ correlation. Fortunately enough, as the latter correlation only relies on prompt 
emission information, we should expect to collect few tens of GRBs with a low redshift dispersion 
in a few months, provided that an adequate $\gamma$-ray instrument acquires the relevant prompt 
emission information, namely the light curve and a broad band spectrum.

While waiting for a sample of calibrators, adequate statistical approaches should be used in 
order to optimally recover cosmographic information from the cosmology-dependent points in the 
Hubble diagram.  The Bayesian formalism presented in \citet{fgga05} is currently the most suitable 
method for this purpose and we apply it here for constraining cosmological parameters by using 
the \relL\ relation. The basic idea of such formalism is to find the best-fitted correlation on 
each point $\bar\Omega$ of the explored cosmological parameter space [for instance 
$\bar{\Omega} = (\bar{\Omega}_{\rm m},\bar{\Omega}_{\Lambda})$] and to estimate, using such a 
correlation, the scatter $\chi^2(\Omega,\bar{\Omega})$ on the Hubble diagram for any given cosmology 
$\Omega$. The conditional probability $P(\Omega|\bar{\Omega})$, inferred from the 
$\chi^2(\Omega,\bar{\Omega})$ statistics, provides the probability for each $\Omega$ given a possible 
$\bar{\Omega}$-defined correlation. By defining $P^\prime(\bar\Omega)$ as an arbitrary probability 
for each $\bar\Omega$--defined correlation, the total probability of each $\Omega$, using the Bayes 
formalism, is given by
\begin{equation}
P(\Omega) = \int P(\Omega|\bar{\Omega})P^\prime(\bar{\Omega})d\bar{\Omega},
\label{bayes}
\end{equation}
where the integral is extended over the available $\bar{\Omega}$ space. Note that from the 
observations one obtains a correlation for each cosmology. Therefore, $P'(\bar{\Omega})$ is actually 
the probability of the given cosmology. Consequently such probability is obtained by putting 
$P'(\Omega)=P(\Omega)$ and solving the integral Eq.~(\ref{bayes}). It should be noted that in the 
conventional use of the Bayes approach, $P'$ is handled as a given prior probability. Here, instead, 
$P'$ and $P$ are just the same probability which is solution of Eq.~(\ref{bayes}).

An elegant Monte Carlo approach allows us to solve Eq.~(\ref{bayes}), i.e. to find the probability 
$P(\Omega)$ from this integral equation. We start by determining the empirical correlation for an 
arbitrary cosmology $\Omega_0$. The $\Omega_0$-defined correlation is used to calculate on the Hubble 
diagram the probability distribution $p_0(\Omega|\Omega_0) \propto exp(-\chi^2(\Omega,\Omega_0)/2)$. 
From this probability, we randomly draw the cosmology $\Omega_1$, which is used to determine again 
the empirical correlation. Then, with the $\Omega_1$-defined correlation, we calculate a new probability 
distribution $p_1(\Omega)$ that is averaged with $p_0(\Omega)$. The result is the probability 
distribution $P_1(\Omega)$. From this probability, a cosmology $\Omega_2$ is randomly drawn and is 
used to determine again the empirical correlation. Applying this correlation in the Hubble diagram 
gives a new probability distribution $p_2(\Omega)$. The new probability $p_2(\Omega)$ is averaged 
with the previous ones, $p_0(\Omega)$ and  $p_1(\Omega)$, giving the probability distribution 
$P_2(\Omega)$. The cycle is repeated until convergence, i.e. until $P_i(\Omega)$ did not change 
significantly with respect to $P_{i-1}(\Omega)$. The convergence should be fast if the empirical 
correlation is not too sensitive to cosmology. On the contrary, if the correlation is strongly 
dependent on cosmology, then convergence cannot be attained with this method. Numerical experiments 
show that for the \relL\ relation, convergence is attained after a few thousands of cycles with a 
result that is independent of the choice of the initial cosmology $\Omega_0$. By introducing some 
numerical techniques, the convergence can be attained after hundreds of cycles.

The described formalism is very different from assuming that the correlation is known ({\it unique 
and well calibrated}) as done in the previous section. It is also different from scanning {\it directly} 
the $\chi^2$ parameter for all points in the $\Omega$ plane by minimizing the scatter of the data 
points around a correlation that is found in the very same $\Omega$ point, and therefore changes 
from point to point, as we did in \S~3. Besides, the Bayesian formalism is less affected than the 
{\it direct} method by possible discontinuities, like the loitering line in the 
$({\Omega}_{\rm m},{\Omega}_{\Lambda})$ plane, and it avoids spurious divergences. With the Bayesian 
formalism, the observational information, provided by correlations like the \relL\ one, is optimally 
extracted for cosmographic purposes as long as this kind of correlations remain uncalibrated at low $z$.

It should be remarked that there is no formal and rigorous mathematical method for solving the 
``circularity problem''. However, a comparison of Figures~1 and 8 clearly shows how the constraints 
improve when one or another formalism is used. Xu et al. (2005)  have also shown the much better 
performance of the Bayesian formalism as compared to the other methods. 

Finally, we mention that the Bayesian approach has been also used by us \citep{Firmani06b} to constrain 
cosmological parameters from the Supernova Legacy Survey \citep[SNLS,][]{Astier06}. The SNLS data are 
given in such a way that they also require a cosmology-dependent calibration. We have found constraints 
and CL contours in the $(\Omega_m,\Omega_\Lambda)$ plane similar to those of Astier et al. (2006), who 
used the direct $\chi^2$ minimization method; if anything, our CL contours were slightly narrower. This 
shows that our method is working for what it has been designed, namely to optimize the data from 
quasi-standard candles in the Hubble diagram.

\section{Results}

In this section we present the results on the cosmological constraints obtained with the Bayesian 
formalism (\S~4.3) applied to the tight \relL\ correlation defined with the 19 long GRBs distributed 
in redshift to up to $4.5$. Following \S~4.1, we proceed to constrain only 2 parameters each time. 
In all the models we fix $h=0.71$. It should be emphasized that our results represent a first attempt, 
still using a dataset with few numbers and with a quality of the observational information not yet at 
the level of SNIa. However, these results allow us to quantify the potentiality of the GRB \relL\ 
correlation as an independent cosmological tool.
 
For comparison purposes, we will also show the cosmological constraints provided by the SNIa 
``gold set'' \citep[][$z<1.67$]{Riess04}. The latter were derived by using the standard (direct) 
\chs-fitting procedure. It is worth to mention that the results on cosmological constraints 
recently presented by the ``Supernova Legacy Survey'' group \citep[$z<1.01$;][]{Astier06} show 
distinct trends that are not shared by the ``gold'' set \citep[see also][]{NP05,Firmani06b}.


\begin{figure}[!t]
\vspace{0.1cm}
\centerline{\includegraphics[width=0.98\columnwidth]{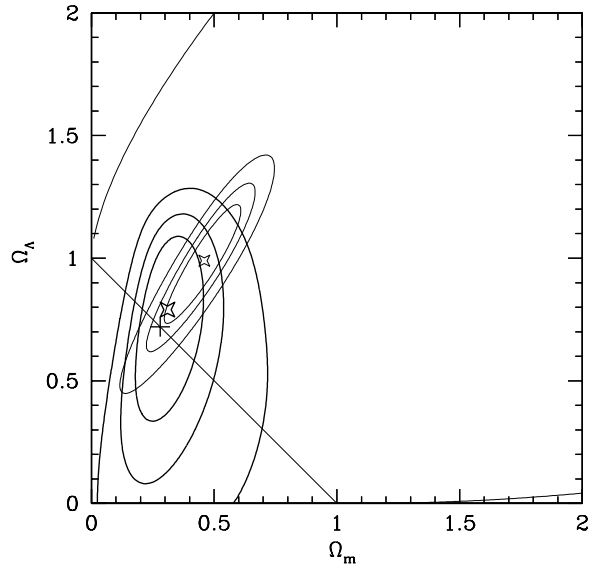}} 
\caption{Constraints on the (\Om, \OL) plane for a $\Lambda$ cosmology 
from the GRB Hubble diagram using our Bayesian method to circumvent the 
circularity problem (thick-line ellipses) and from the ``gold set'' SNIa 
Hubble diagram (thin-line ellipses). The ellipses are contours at 68.3\%, 
95.5\%, and 99.7\% CL's.  The other curves in the plot are as in Figure~1.}
\label{fig8} 
\vspace{0.4cm}
\end{figure}


\subsection{$\Lambda$ Cosmology}

In Figure~\ref{fig8} we show the 1$\sigma$, 2$\sigma$, and 3$\sigma$ CL's (thick lines) for 
the \Om\ and \OL\ parameters.  Notice how the CL's improve with respect to those obtained 
with the simplest direct $\chi^2$ minimization method used in \S~3 (Figure~\ref{fig1}). The 
best-fit cosmology (see the star symbol in Figure~\ref{fig8}) corresponds to \Om=$0.31^{+0.09}_{-0.08}$, 
\OL=$0.80^{+0.20}_{-0.30}$ (1$\sigma$ uncertainty). This result is very close to the flat 
geometry case. The concordance model is well within the 1$\sigma$ CL. If the flat geometry 
case is assumed (i.e. \Ot = 1), our statistical analysis constrains $\Om = 0.29^{+0.08}_{-0.06}$.

The constraints on the $\Lambda$ cosmology parameters that we obtain with GRBs {\it alone} 
are consistent with those obtained through several other cosmological probes 
\citep[e.g.][]{Hawkins03,schu03}. In turn, this result gives us confidence that GRBs can be 
used as cosmological probes.

In Figure \ref{fig8} are also shown the best-fit values (star symbol) and CL regions (thin 
lines) that we obtain with the SNIa ``gold set'' \citep[][]{Riess04}. As these and other 
authors (e.g. Choudhury \& Padmanabhan 2004; Jassal et al. 2005; Nesseris \& Perivolaropoulos 
2005b) have shown, the ``gold set'' provides constraints on \Om\ and \OL\ that are only 
marginally consistent with the concordance model or the {\it WMAP} CBR constraints.


\begin{figure}[!t]
\vspace{0.1cm}
\centerline{\includegraphics[width=0.98\columnwidth]{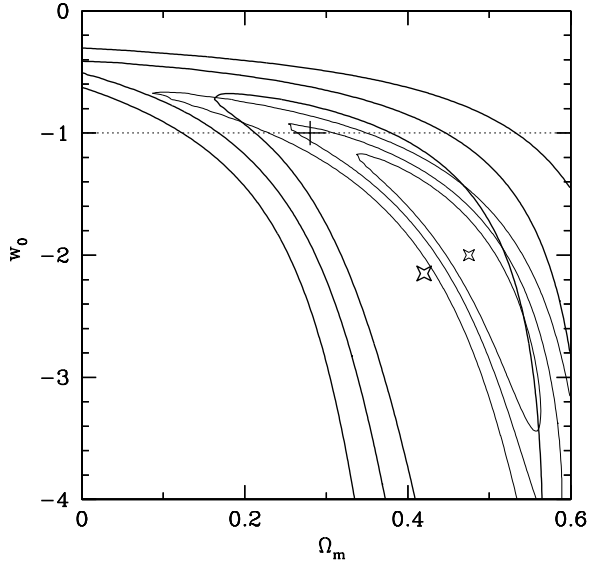}}
\caption{ Constraints on the (\Om, $w$) plane for a flat cosmology with 
static DE from the GRB Hubble diagram using the Bayesian formalism to 
solve optimally the circularity problem (thick-line ellipses) and from 
the ``gold set'' SNIa Hubble diagram (thin-line ellipses). The bent 
ellipses are contours at 68.3\%, 95.5\%, and 99.7\% CL's.}
\label{fig9}
\vspace{0.4cm}
\end{figure}



\begin{figure}[!t]
\vspace{0.06cm}
\centerline{\includegraphics[width=0.96\columnwidth]{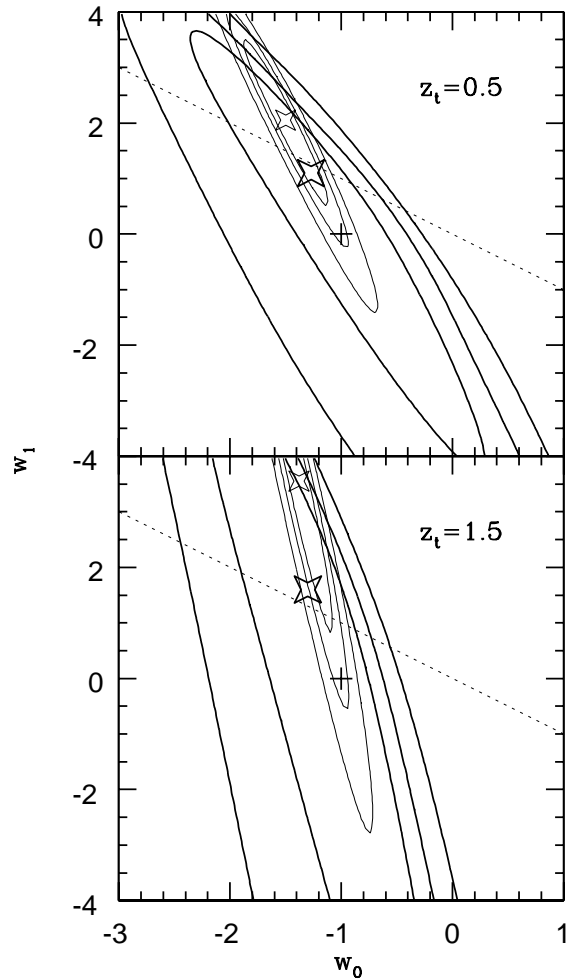}} 
\caption{Constraints on the (\wo,\wu) plane for a flat cosmology with 
dynamic DE and \Om=0.28 from the GRB Hubble diagram using the Bayesian 
formalism to solve optimally the circularity problem (thick-line ellipses) 
and from the ``gold set'' SNIa Hubble diagram (thin-line ellipses). Upper 
and lower panels are for \zt=0.5 and 1.5, respectively. The ellipses are 
contours at 68.3\%, 95.5\%, and 99.7\% CL's. The diagonal dot-dashed line 
is the upper limit in the (\wo,\wu)-plane allowed by CMB constraints.}
\label{fig10}
\vspace{0.45cm}
\end{figure}


\subsection{Flat Cosmology with Static ($w=$const) DE}

\vspace{0.05cm}

Figure~\ref{fig9} shows the 1$\sigma$, 2$\sigma$, and 3$\sigma$ CL regions on the (\Om, $w$) 
plane for models with static DE and flat geometry, using the GRB sample (thick lines) and the 
SNIa ``gold set'' (thin lines).

\vspace{0.15cm}

The degeneracy here is relevant and higher than the corresponding degeneracy seen in Figure~\ref{fig8}. 
This feature is consistent with the discussion of Figures~\ref{fig5} and \ref{fig6} of the previous 
section and has to do with the small rotation of constant \dl\ lines for different redshifts. The 
reduction of such a degeneracy will be possible by reducing the GRB observational uncertainty as 
well as by increasing the number of objects \citep[see e.g.][for a simulation]{Ghirla05}.

\vspace{0.15cm}

The $\Lambda$ case ($w=-1$) is consistent at the 68.3$\%$ CL with the GRB constraints, for values of 
$\Om = 0.29^{+0.08}_{-0.07}$. The concordance model is well inside the 68.3$\%$ CL. For a prior 
\Om=0.28, we obtain $w= -1.07^{+0.25}_{-0.38}$. Note that the $\Lambda$ case is not consistent with 
the SNIa ``gold set'' at the 68.3\% CL.

\subsection{Flat Cosmology with Dynamical DE}

\vspace{0.1cm}

Formal constraints on (\wo, \wu) (see Eq.~\ref{w}), assuming flat-geometry cosmologies (and $\Om=0.28$) 
with dynamical DE, are presented in Figure~\ref{fig10}. Upper and lower panels refer to $\zt=0.5$ and 
$\zt=1.5$, respectively. The thick and the thin line ellipses are the 1$\sigma$, 2$\sigma$ and 3$\sigma$ 
CL regions for the GRB and SNIa ``gold'' samples, respectively. The $\Lambda$ case (\wo=$-1$ and \wu=0, 
which reduces to the concordance model because of the assumption that \Om=0.28), is within the 
1$\sigma$ CL.

The typical redshift of our GRB sample is z$\approx$1.5. The best information provided by GRBs 
on the value of $w(z)$ is expected at the same redshift. Our analysis for $\zt=0.5$ gives 
$w(1.5)=-0.5^{+0.7}_{-1.0}$, and for $\zt=1.5$ it gives $w(1.5)=-0.5^{+0.9}_{-1.9}$. The (still) 
large uncertainties in the data and the small number of objects do not allow to constrain \zt\ 
as a third parameter.

Again, note that the constraints provided by the SNIa ``gold set'' are not consistent with the 
concordance model values of \wo\ and \wu\ at the 68.3\% CL.

In general, the ``gold'' SNIa constraints tend to favor low values of \wo\ and large values of \wu, 
implying (i) a strong evolution of $w(z)$ in the range $0 \lsim z \lsim 1$, and (ii) a significant 
probability for $w(z)$ to cross the $w=-1$ line (phantom divide line, see also e.g. Riess et al. 
2004; Alam et al. 2004; Nesseris \& Perivolaropoulos 2005b). As reviewed by the last authors, if 
observations show a significant probability for $w(z)$ to cross the phantom divide line, then all 
minimally coupled single scalar field models would be ruled out as DE candidates, leaving only 
models based on extended gravity theories and combinations of multiple fields. It is therefore a 
key observational task to determine whether $w(z)$ crosses the $w=-1$ line or not.  The SNIa ``gold 
set'' rejects models that avoid the phantom dividing line at the 1$\sigma$ CL, while our results 
with GRBs allow these models (including the concordance one) at the 1$\sigma$ CL, though the 
uncertainties for the latter are still much larger than for the former.

Finally, we should emphasize that Eq.~(\ref{w}) is just a mathematical parametrization for the 
evolution of $w$, but not a physical model of DE. Although Eq.~(\ref{w}) describes the evolution 
of $w$ up to any arbitrary large $z$ once its parameters are determined, the changes in $w$ with 
$z$ suggested by the observational constraints are formally valid only within the redshift range 
of the observational data. For example, the constraints shown in Figure~\ref{fig10} cannot be 
used to extrapolate the behavior of $w(z)$ as given by Eq.~(\ref{w}) to $z'$s higher than $\sim 3$, 
and $\sim 1$ for the GRB and SNIa data, respectively.  

In fact, at high redshifts there are several observational limits to the values of the parameters 
of Eq.~(\ref{w}). The most important is related to the CMB anisotropies. The CMB data require 
$\Omega_{\rm DE} \lsim 0.1$ at the redshift of recombination, $z=1100$ \citep{Caldwell04}. For the 
``Rapetti'' parametrization that we are using (Eq.~[\ref{w}]) and assuming flat geometry, this 
condition implies that $\wo + 0.86\wu \lsim -0.095$, which is close to the general upper limit of 
$\wi = \wu + \wo \lsim 0$ found in the analysis of \textit{WMAP} and other data sets by \citet{raw05}. 
The dashed line in Figure~\ref{fig10} corresponds to this limit. Interestingly enough, the 
best-fitting point from the GRB sample in the (\wo, \wu) plane obeys the CMB constraint for $\zt=0.5$, 
being slightly out of this constraint for $\zt=1.5$. Instead, for the SNIa ``gold set'' the 
best-fitting points in both cases are far away from the CMB constraint.

\section{Summary and Discussion}

Firmani et al. (Paper I) found a very tight correlation among three GRB quantities in their rest 
frame, \Liso, \Ep\ and \tdur. These quantities were calculated from the $\gamma-$ray prompt emission 
spectra and light curve, without the addition of any quantity derived from the afterglow, apart 
from the redshift.

Here we have used this tight correlation to ``standardize'' the energetics of the currently available 
sample of 19 GRBs, and to construct an observational Hubble diagram up to the record redshift of 
$z=4.5$ and independent from SNIa. Based on the behavior of the luminosity distance as a function of 
different cosmological parameters (\S~4.2), we have pointed out that samples of standard candles 
distributed over a wide redshift range are strongly desired for breaking the degeneracy of the 
cosmological parameters. To overcome the circularity problem that arises due to the lack of a local 
cosmology-independent calibration of the \relL\ relation, we have applied a Bayesian formalism 
developed in \citet{fgga05} and further discussed here. The main results on the cosmological 
constraints are:
\begin{itemize}
\item The \relL\ correlation is sensitive to the cosmological parameters of the $\Lambda$ cosmology 
(\S~3), having a minimum $\chsr$ in (\Om, \OL) = (0.31, 0.78), very close to the concordance model 
(Figure~1).
\item For the $\Lambda$ cosmology, using the Bayesian formalism, the best-fitting values for \Om\ 
and \OL\ are $0.31^{+0.09}_{-0.08}$ and $0.67^{+0.20}_{-0.30}$ (1$\sigma$ uncertainty), respectively. 
This result is very close to the flat geometry (Figure~\ref{fig8}). The $\Lambda$CDM concordance 
model (\Om =0.28 and \OL = 0.72) is well within the 68.3\% CL. If one assumes flat geometry, then 
we find $\Om = 0.29^{+0.08}_{-0.06}$.
\item For constant $w$ models (static DE) with flat geometry, the $\Lambda$ case ($w=-1$) is 
consistent at the 68.3$\%$ CL for values of $\Om\ = 0.29^{+0.08}_{-0.07}$. The $\Lambda$CDM 
concordance model is still within the 68.3\% CL.
\item For models with dynamical DE, we have parametrized $w(z)$ according to Eq.~(\ref{w}) and 
used $\zt = 0.5$ and $\zt = 1.5$.  Assuming a flat geometry and \Om=0.28, the $\Lambda$ case 
(\wo=$-1$ and \wu=0, which also in this case corresponds to the concordance model) is again 
within the 68.3$\%$ CL.  Interestingly enough, the constraint that the CMB data ($z=1100$) provide 
on $w(z)$ as given by Eq.~(\ref{w}) ($\wo+0.86\wu\ \lsim-0.095$), is consistent with the constraints 
found with GRBs.
\end{itemize}

We conclude that the different constraints provided by the GRB sample are consistent at the 
68.3\% CL with the $\Lambda$CDM concordance model. This is not the case of the SNIa ``gold 
set''.  Also, the GRB constraints for flat-geometry models, with DE equation of state parameter 
either constant or varying with $z$, are consistent with the constant $w(z)$ case at the 68.3\% 
CL, while the ``gold set'' SNIe are not. These results show that the GRB method presented here 
offers already a competitive and reliable way to discriminate cosmological parameters.

The use of the correlation \relL\ among prompt $\gamma$-ray quantities has proved to be a 
promising, model-independent and assumption-free method for constructing the observational 
Hubble diagram up to high redshifts. The accuracy that this correlation provides in constraining 
cosmological parameters with the current available set of useful GRBs is better than that 
found with other correlations (either the ``Liang \& Zhang'' correlation or the ``Ghirlanda'' 
correlation). Most importantly, the advantage of the \relL\ correlation is that it does not 
involve any quantity related to the afterglow.

Compared to SNIa, the GRB cosmological constraints are less accurate. This is due to the 
still low number of GRBs having the required data as well as to the relatively large 
uncertainties associated with these data. However, GRBs provide valuable complementary 
cosmographic information, in particular due to the fact that GRBs span a much wider redshift 
distribution than SNIa. As discussed in \S~4.2, some degeneracies appear when constraining 
the cosmological parameters with samples of ``standard candles'' limited only to low redshifts. 
The results presented in this paper are a clear proof of the potentiality of using the GRB 
\relL\ relation for cosmographic purposes. After the completion of this paper, a paper by 
Schaefer (2007), where a combination of several (noisy) empirical correlations of GRBs was 
used to construct the Hubble diagram up to $z=6.4$, appeared posted in the arXiv preprint 
database service. The constraints on the cosmological parameters obtained in that work are 
similar to ours, though the data and methodology are very different from the ones presented here.  

\vspace{0.08cm}

It is worth to mention that as more data of higher quality appear, some assumptions made 
concerning the cosmographic use of the \relL\ relations will either be accepted or refused. 
For example, in order that this relation be useful for cosmography, it should not evolve, 
or the way in which it changes with $z$ should be known. It is also important to improve 
the quality of the data in order to reduce the scatter, as well as to increase the number 
of usable GRBs. So far, the best physical justification of the \relL\ relation derives from 
its scalar nature, which explains its reduced scatter because it is independent of the 
relativistic factor $\Gamma$ (Paper I). A full physical interpretation of this relation is 
highly desirable, in particular to avoid any uncertainty concerning observational selection 
effects\footnote{In a recent paper, appeared after the first submission of the present one, 
Thompson, Rees \& Meszaros (2006) have suggested some interesting hints to understand the 
origin of the \relL\ relation.}. However, as an empirical relation used like a distance 
indicator tool, the main concern is related to reducing the observational scatter. Just 
recall the case of the famous Tully-Fisher relation for disk galaxies, which has been used 
as distance indicator for more than twenty years, though until recently its physical 
foundation was not clear.

\vspace{0.08cm}

Another potential problem for high-redshift GRBs as a cosmological tool is gravitational 
lensing which systematically brightens distant objects through the magnification bias and 
increases the dispersion of distance measurements \citep[e.g.][]{Porciani00,Oguri06}. Recently, 
\citet{Oguri06} simulated the gravitational lensing effects on the Hubble diagram constructed 
with {\it Swift}-like GRBs following a reasonable luminosity function \citep{fagt04}. They 
showed that lensing bias is not drastic enough to change constraints on dark energy and its 
evolution. However, they emphasized that the amount of the bias is quite sensitive to the 
shape of the GRB luminosity function. Thus, an accurate measurement of the luminosity function 
is important in order to remove the effect of gravitational lensing and to obtain unbiased 
Hubble diagram.

\vspace{0.08cm}

We finish by emphasizing that the ideal strategy to follow in the future is to combine the 
SNIa and GRB data sets, and to adopt the same methods of handling these data sets of ``standard 
candles'' in order to construct their joint Hubble diagram, and thus constrain the cosmological 
parameters. Of course, the dominant information will be that of SNIa (they outnumber GRBs and 
the uncertainties on their luminosity are smaller than for GRBs), but GRBs provide valuable 
information at high redshifts which helps to partially overcome parameter degeneracies and 
biases. This program is carried out elsewhere \citep{Firmani06b}.

\vspace{0.1cm}

\acknowledgements

We thank Giuseppe Malaspina for technical support, and Jana Benda for grammar corrections. 
We are grateful to the anonymous referee for a through and constructive report that helped 
to improve the content of the paper. VA-R. gratefully acknowledges the hospitality extended 
by Osservatorio Astronomico di Brera. This work was supported by PAPIIT-UNAM grant IN107706-3 
and by the Italian INAF MIUR (Cofin grant 2003020775\_002).

\end{document}